\UseRawInputEncoding
\documentclass[twoside,twocolumn]{article}
\usepackage{fitee}
\usepackage[colorlinks, linkcolor=black, anchorcolor=black, citecolor=black, breaklinks = true]{hyperref}		
\pdfoutput=1
\usepackage{epstopdf}
\usepackage{cite}
\usepackage{amsmath,amssymb,amsfonts}
\usepackage{algorithmic}
\usepackage{graphicx}
\usepackage{subfigure}
\usepackage{textcomp}
\usepackage{xcolor}
\usepackage{gensymb}
\usepackage{multirow}
\usepackage{booktabs}  
\usepackage{threeparttable}
\usepackage{amssymb}
\usepackage{makecell}
\usepackage{upgreek}
\usepackage{bm}
\addto{\captionsenglish}{%
  
}
\begin{document}
\title{Frequency-Angle Two-Dimensional Reflection Coefficient Modeling Based on Terahertz Channel Measurement}

\author[$\dagger$1]{Zhao-wei CHANG}%
\author[$\dagger$$\ddagger$1]{Jian-hua ZHANG}
\author[1]{Pan TANG}
\author[1]{Lei TIAN}
\author[1]{Li YU}
\author[2]{Guang-yi LIU}
\author[2]{Liang XIA}

\affil[1]{State Key Lab of Networking and Switching Technology, Beijing University of Posts and Telecommunications, Beijing 100876, China}
\affil[2]{China Mobile Research Institute, Beijing, 100053, China}

\shortauthor{Chang et al}	
\authmark{}
\corremailA{changzw12345@bupt.edu.cn}
\corremailB{jhzhang@bupt.edu.cn}
\emailmark{$\dagger$}	

\dateinfo{Received mmm.\ dd, 2016;	Revision accepted mmm.\ dd, 2016;    Crosschecked mmm.\ dd, 2017}

\abstract{Terahertz (THz) channel propagation characteristics are vital for the design, evaluation, and optimization for THz communication systems. Moreover, reflection plays a significant role in channel propagation. In this letter, the reflection coefficient of the THz channel is researched based on extensive measurement campaigns. Firstly, we set up the THz channel sounder from 220 to 320 GHz with the incident angle ranging from 10$\degree$ to 80$\degree$. Based on the measured propagation loss, the reflection coefficients of five building materials, i.e., glass, tile, aluminium alloy, board, and plasterboard, are calculated separately for frequencies and incident angles. It is found that the lack of THz relative parameters leads to the Fresnel model of non-metallic materials can not fit the measured data well. Thus, we propose a frequency-angle two-dimensional reflection coefficient model by modifying the Fresnel model with the Lorenz and Drude model. The proposed model characterizes the frequency and incident angle for reflection coefficients and shows low root-mean-square error with the measured data. Generally, these results are useful for modeling THz channels.}

\doi{10.1631/FITEE.1000000}	
\code{A}
\clc{TP}
\publishyear{2018}
\vol{19}
\issue{1}
\pagestart{1}
\pageend{5}
\support{Project supported by the National Science Fund for Distinguished Young Scholars (No. 61925102),  the National Key R\&D Program of China (No. 2020YFB1805002), the National Natural Science Foundation of China (No. 62031019, No. 92167202, No. 62101069), BUPT-CMCC Joint Innovation Center.}

\orcid{Zhao-wei CHANG, http://orcid.org/0000-0002-8689-410X}	
\articleType{Science Letters:}
\maketitle

\section{Introduction} \label{1}
To meet the increasing demand for higher data rates, terahertz (THz) communication between 0.1 to 10 THz attracts a great deal of attention due to its wide bandwidth, which can support much higher speed data rates from tens of Gbps to a few Tbps than millimetre wave (mmWave) communication \citep{r27,r5,r99}. THz channel propagation characteristics play an important role in the design, evaluation, and optimization of THz communication systems \citep{r10}. Furthermore, due to the short wavelength at THz band, the propagation mechanisms, i.e., reflection and diffraction, may change \citep{r3,r7}. Therefore, it is quite necessary to measure and model the reflection coefficients at THz bands.

Recently, a number of measurement campaigns have been conducted for characterizing reflection coefficients at THz bands. For example, reflection coefficients of materials are measured in \citep{r19} with a dielectric lens antenna from 207 to 247 GHz. In \citep{r22}, measurements of reflection coefficients with incident angles of 10$\degree$, 30$\degree$, 60$\degree$, and 80$\degree$ are carried out for drywall at 142 GHz. \citep{r24} presents reflection coefficients THz time-domain spectroscopy (THz-TDS) measurements for different building materials from 70 to 350 GHz at the angle ranging from 20$\degree$ to 75$\degree$. In \citep{r25}, THz-TDS measurements are conducted from 100 GHz to 4 THz at incident angles varying from 35$\degree$ to 70$\degree$. In short, reflection coefficients of different materials have been measured in some THz bands. However, the dependence of reflection coefficients on frequencies, incident angles, and materials, is investigated partially. Also, there is still no comprehensive model that can accurately describe the above-mentioned dependence.

In this letter, we try to fill this gap and present a reflection coefficients measurement of five materials from 220 to 320 GHz with the incident angle from 10$\degree$ to 80$\degree$. Based on the measurement, the dependence of reflection coefficients on frequency, angle, and material is analyzed. Moreover, frequency-angle two-dimensional reflection coefficients (FARC) model is proposed. The contributions of this letter are as follows:
\begin{itemize}
\item We present an extensive reflection coefficients measurement campaign at frequencies ranging from 220 to 320 GHz and the incident angles vary from 10$\degree$ to 80$\degree$. The extensive measurement data are collected with five building materials，, i.e., glass, tile, aluminium alloy, board, and plasterboard.
\item We investigate the dependence of reflection coefficients on frequencies, incident angles, and materials based on the measurement data and Fresnel model, which provides guidance for modeling reflection coefficients.
\item By modifying the Fresnel model with the Lorenz and Drude model, we propose the FARC model at THz bands. Based on the measurement results, we get sets of fitting parameters of the FARC model for different materials. The FARC model can well describe the dependence of the reflection coefficients on the frequency and incident angle.
\end{itemize}

\section{TERAHERTZ RRFLECTION COEFFICIENTS MEASUREMENTS}\label{2}
The measurement of reflection coefficients is conducted by a wideband channel sounder. The details of the channel sounder can be referred to in \citep{r26}. The measurement setup, procedures, and data processing will be introduced in this section.
\subsection{Measurement setup and procedures}\label{22}
\begin{figure}[!t]
\centering
\subfigure[Actual measurement]{\includegraphics[width=3.65cm]{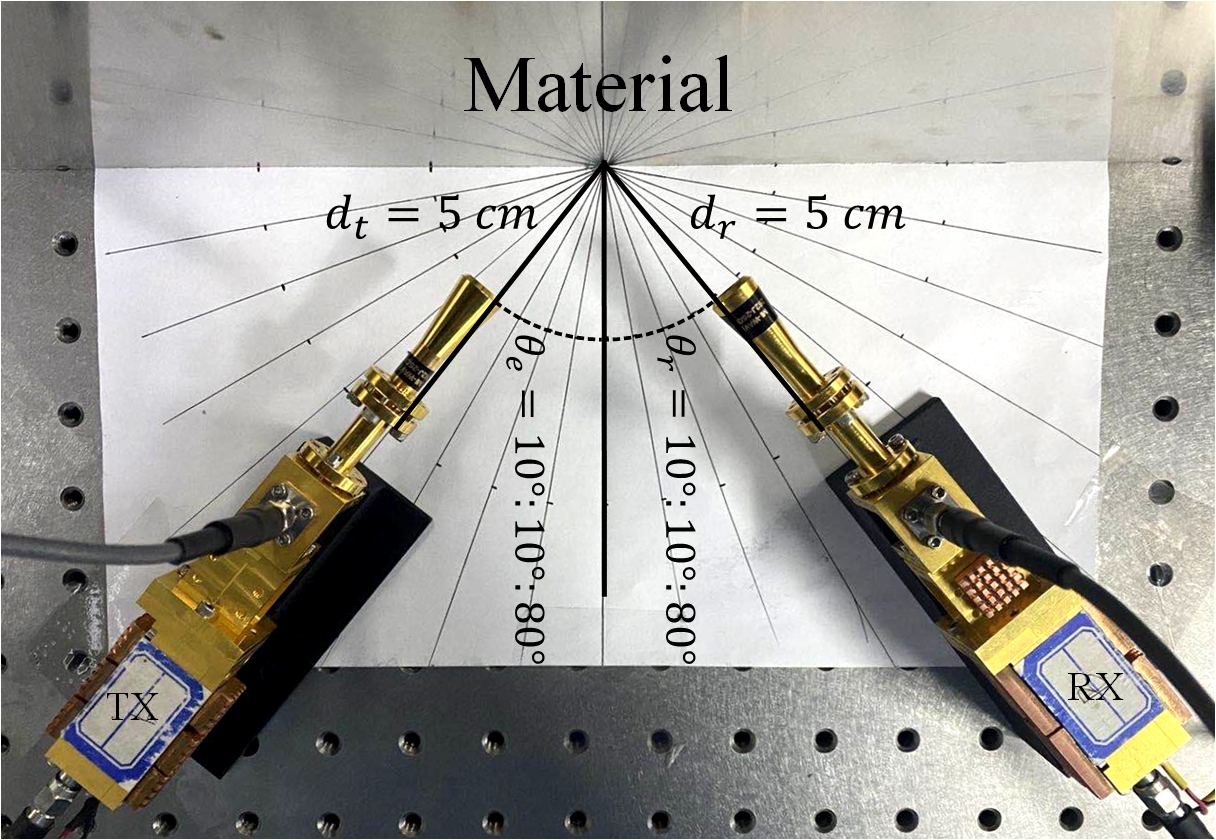}\label{Fig 103}}
\subfigure[Sketch]{\includegraphics[width=4.1cm]{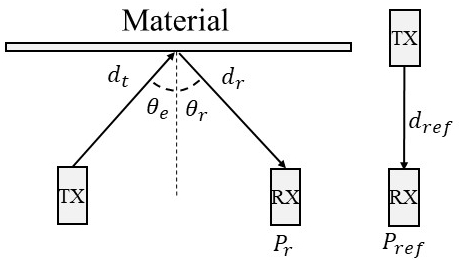}\label{Fig 102}}
\caption{THz reflection coefficients measurement setup.}\label{Fig 4}
\end{figure}

The measurement of reflection coefficients is conducted in the 25 degrees Celsius clean room. Before installing the channel sounder, the system calibration is done after the frequency mixer connected to each other without antennas. The system calibration is introduced in \citep{r26}. After installing the channel sounder, the two antennas are aligned $d_{ref}$ ($d_{ref}$=10 cm) apart to get the reference power $P_{ref}$. Then, the five building materials, i.e., glass, tile, aluminium alloy board, board, and plasterboard, are fixed at the clamp successively under the circumstance of the TX and RX being set as shown in Fig. \ref{Fig 4}a. The distances between TX antenna to the materials $d_t$ and between RX antenna to the materials $d_r$ are both 5 cm. The sketch of measurement setup is shown in Fig. \ref{Fig 4}b. The reflection coefficients of the vertically polarized signal are measured from 10$\degree$ to 80$\degree$ with a 10$\degree$ step. Besides, the measurement is done from 220 to 320 GHz with a 10 GHz step except 270 and 310 GHz. These two frequency points are not measured because the measurement system is unstable when measuring. In each measurement, 50 snapshots of in-phase and quadrature (IQ) data are collected.

\subsection{Data processing}\label{31}
The formula calculating the reflection coefficient can be expressed as \citep{r28}:

\begin{equation}\label{P_R}
\begin{split}
\lvert \Gamma \rvert=\frac{d_t+d_r}{d_{ref}}\sqrt{\frac{P_r}{P_{ref}}},
\end{split}
\end{equation}

\noindent where $P_r$ is the received power by reflected. $P_r$ and $P_{ref}$ are calibrated due to the systematic deviation.
\section{ANALYSIS OF REFLECTION CHARACTERISTICS}\label{3}
This section shows the results of reflection coefficients of five materials. Firstly, frequency and angle dependence of reflection coefficients is investigated by comparing measured results with theoretical modeling. Then, the relationship between reflection coefficients and different materials is described by presenting comparisons among reflection coefficients of five materials.
\subsection{Frequency and angle dependence of reflection coefficients}\label{31}
\begin{figure*}[!t]
\centering
\subfigure[Glass]{\includegraphics[width=5.4cm]{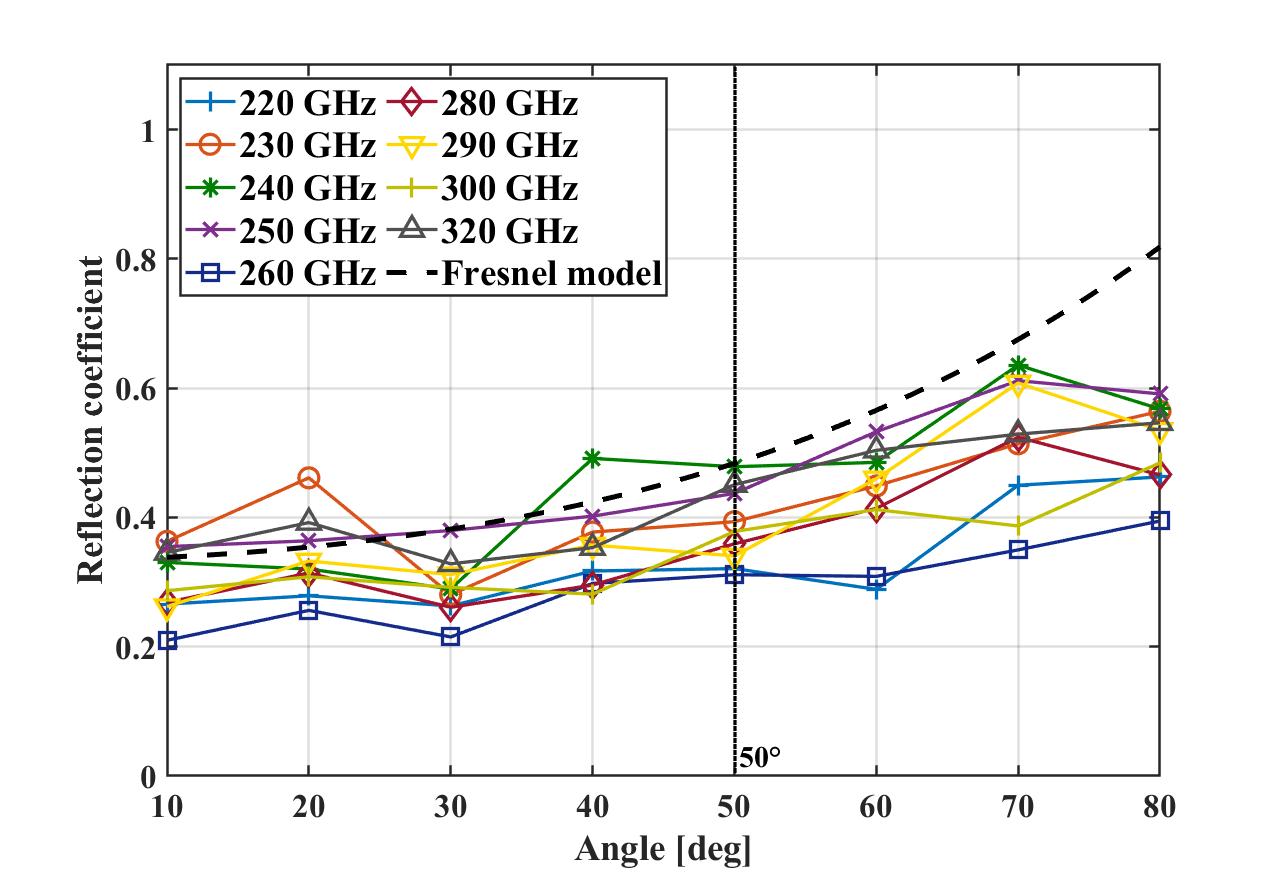}\label{Fig 12}}
\subfigure[Tile]{\includegraphics[width=5.4cm]{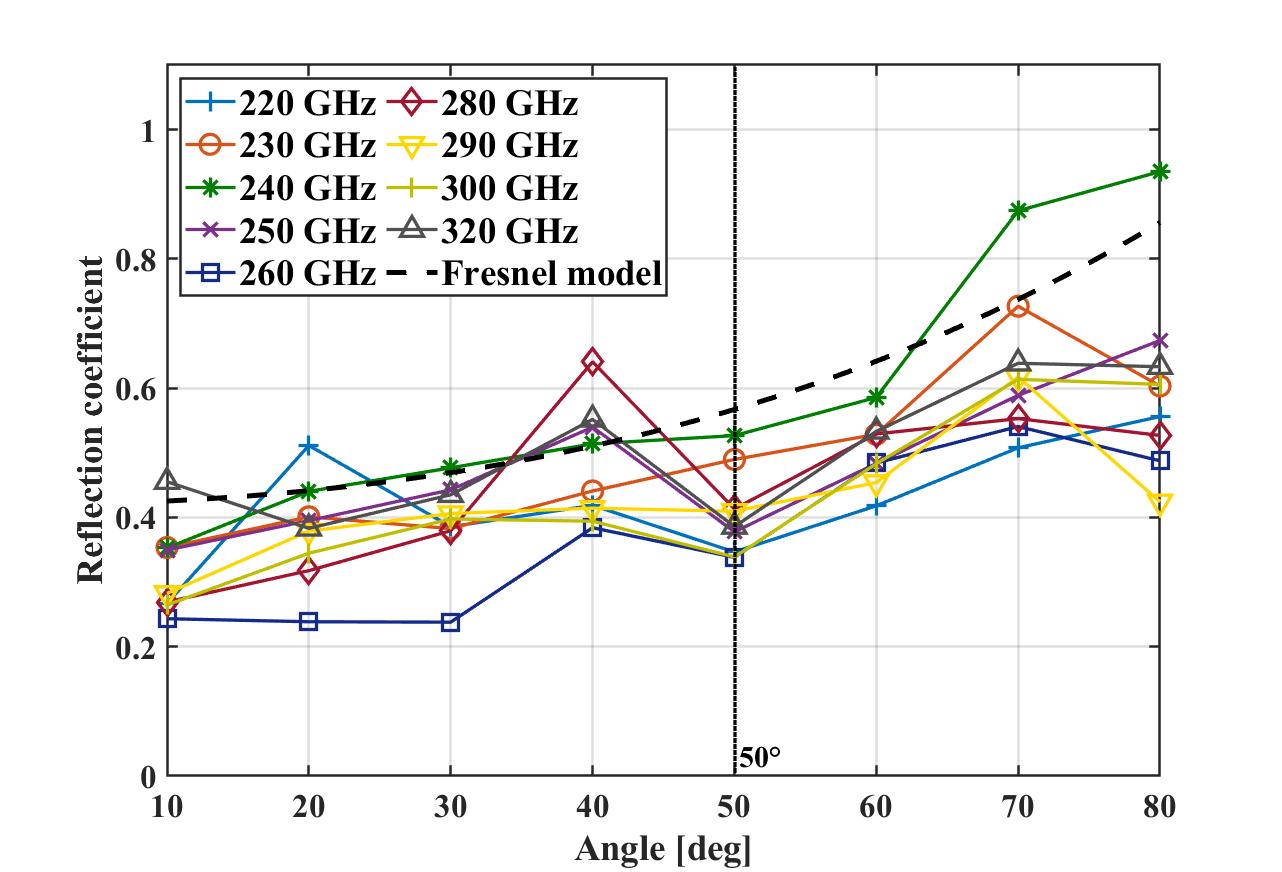}\label{Fig 11}}
\subfigure[Board]{\includegraphics[width=5.4cm]{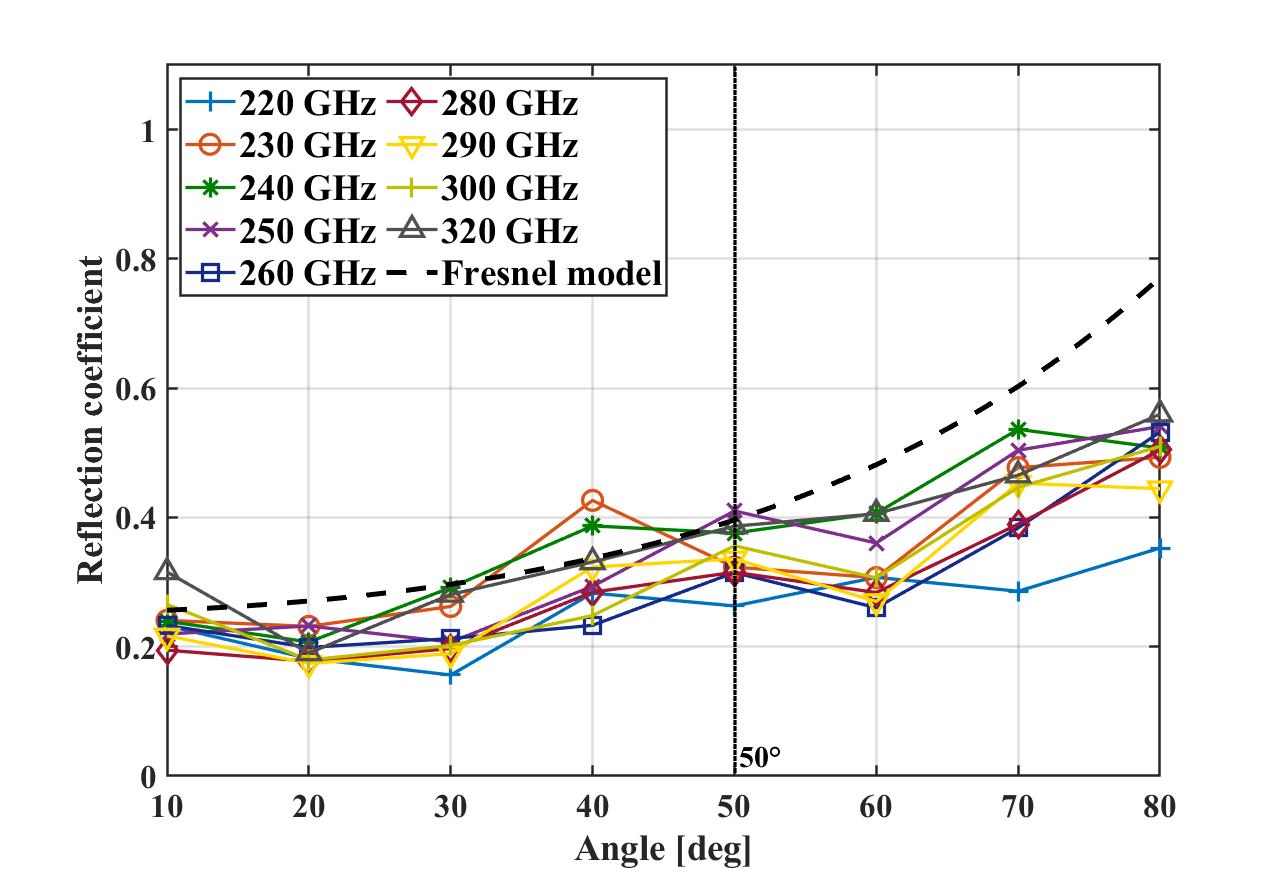}\label{Fig 10}}
\quad
\subfigure[Plasterboard]{\includegraphics[width=5.4cm]{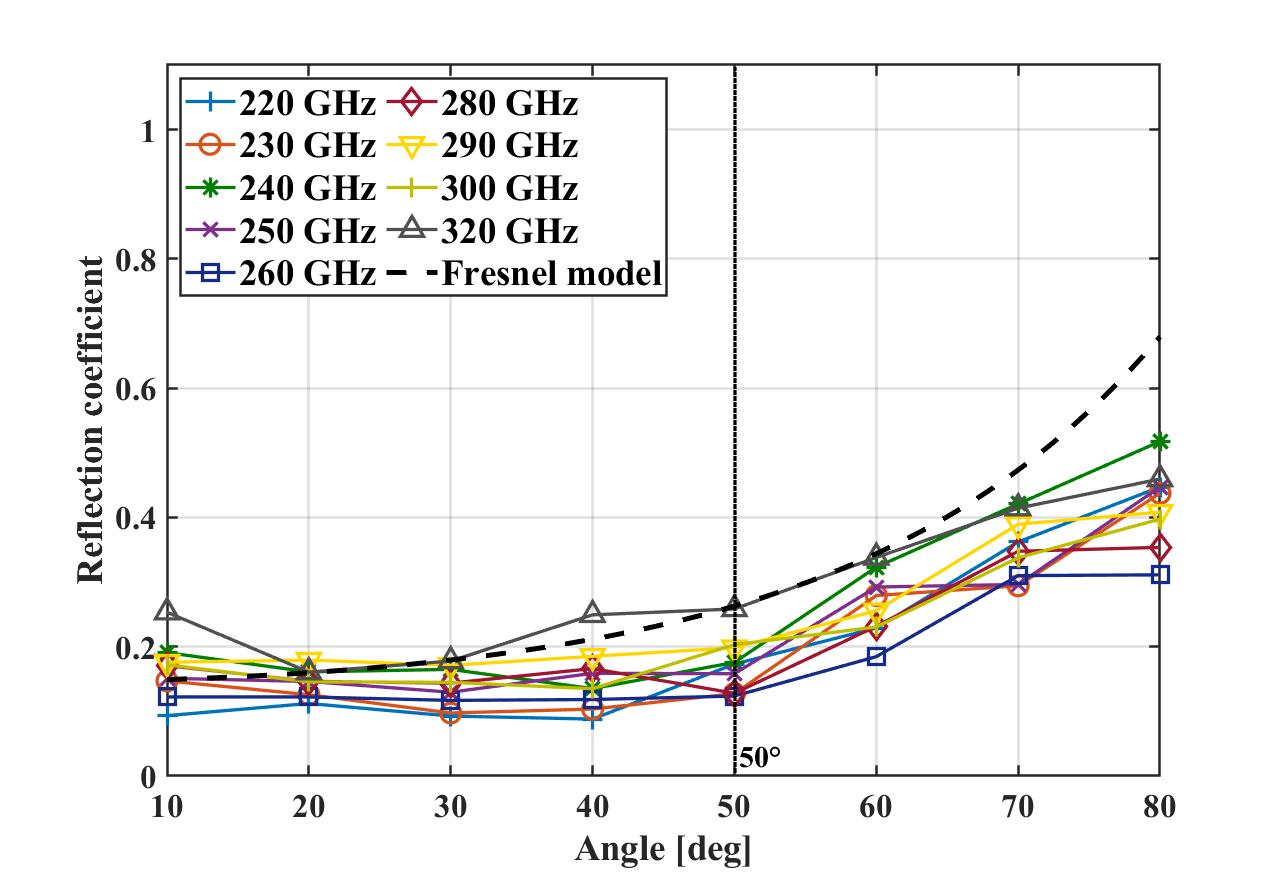}\label{Fig 9}}
\subfigure[Aluminium alloy]{\includegraphics[width=5.4cm]{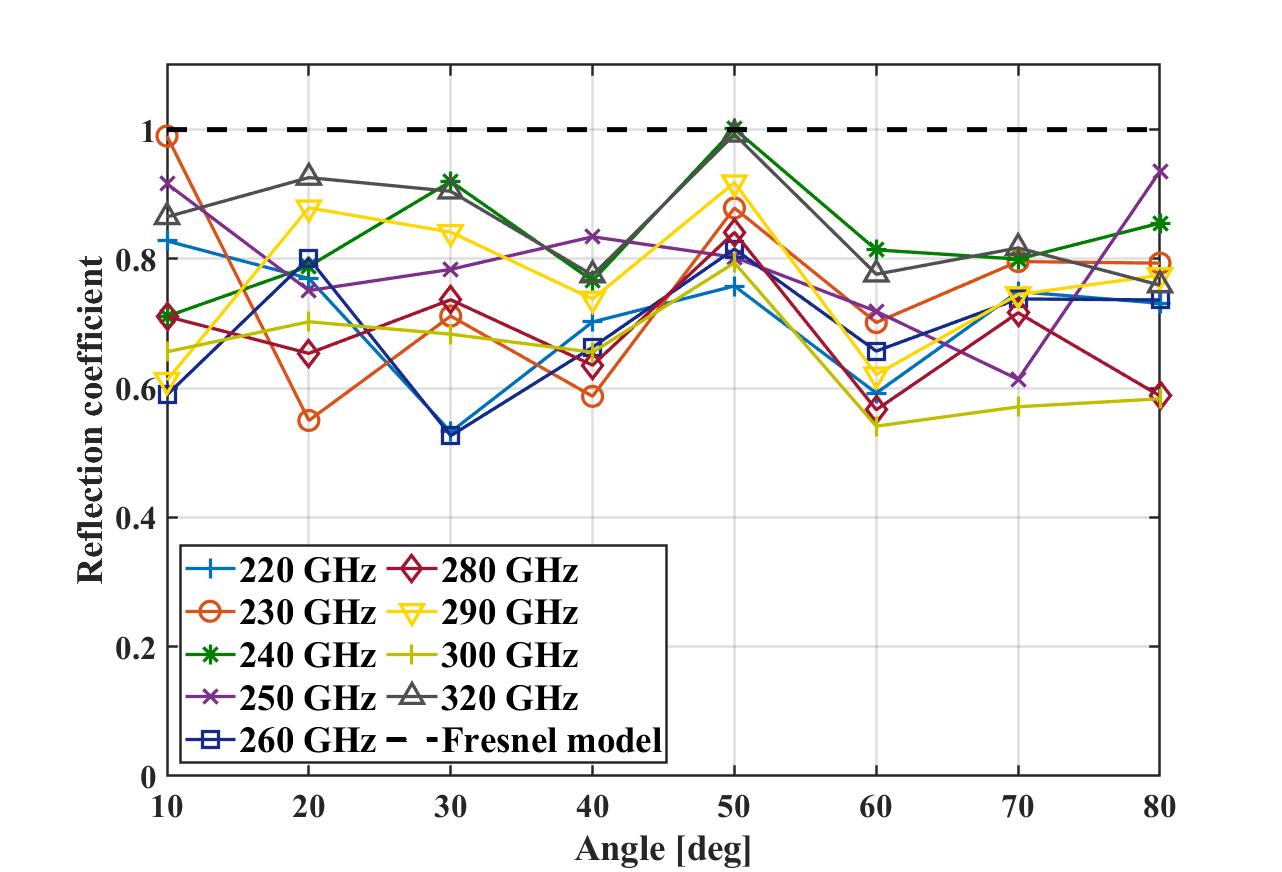}\label{Fig 8}}
\subfigure[Average reflection coefficients]{\includegraphics[width=5.4cm]{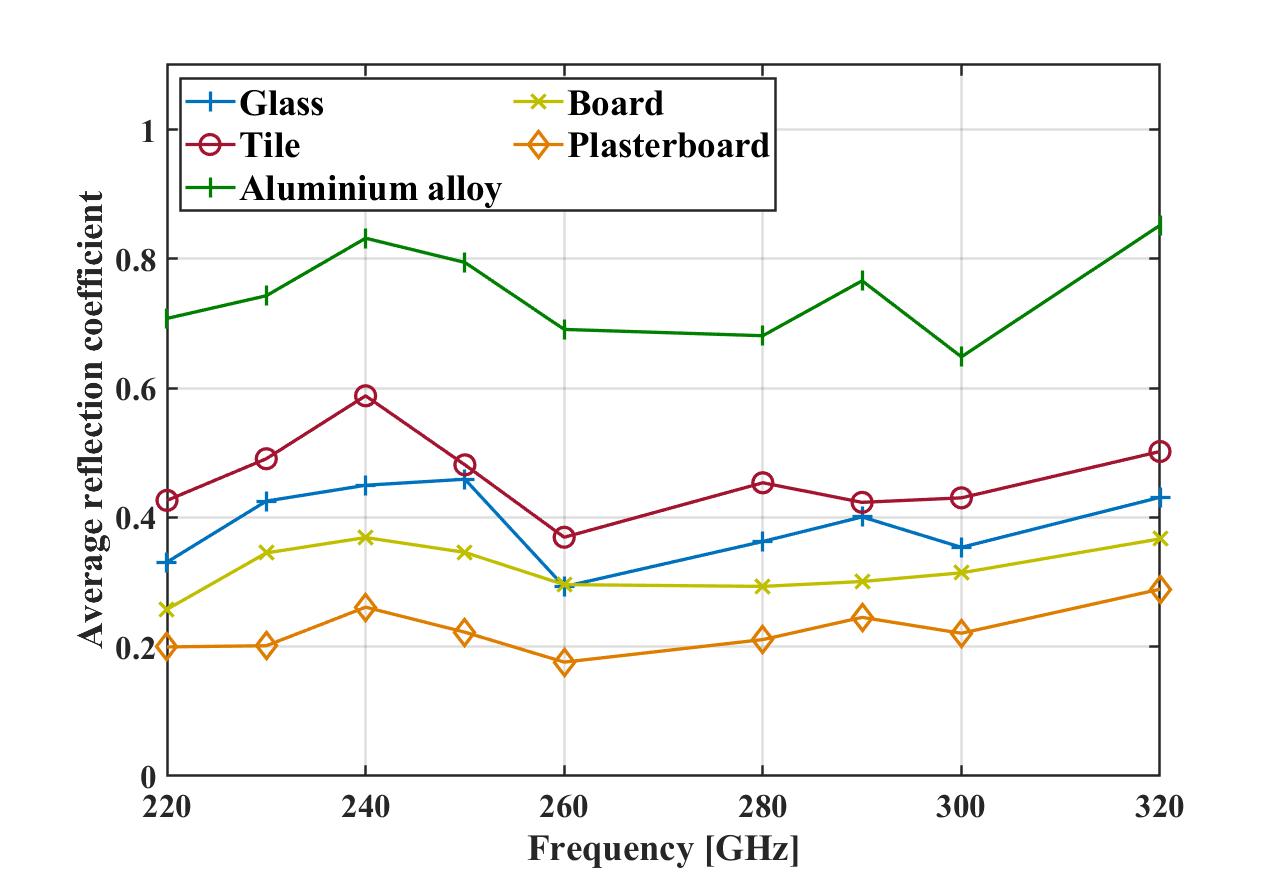}\label{Fig 7}}
\caption{Reflection coefficients for glass, tile, board, plasterboard, aluminium alloy at nine frequency points: measurement results compared with Fresnel reflection coefficients model, and average reflection coefficients of five materials.}\label{Fig 6}
\end{figure*}

To investigate the frequency and the angle dependence of reflection coefficients, the reflection coefficients of five materials are plotted from 10$\degree$ to 80$\degree$ at different frequency points in Fig. \ref{Fig 6}. The results at nine frequency points are shown. As a comparison, the Fresnel reflection coefficients model is given as \citep{r28}:

\begin{equation}\label{e_10}
\begin{split}
\Gamma=e^{-8(\frac{{\uppi}{\upsigma}{\rm{cos}}\theta_e}{\uplambda})^2}\left(\frac{{\rm{cos}}\theta_e-\sqrt{\delta-({\rm{sin}}\theta_e)^2}}{{\rm{cos}}\theta_e+\sqrt{\delta-({\rm{sin}}\theta_e)^2}}\right),
\end{split}
\end{equation}

\noindent where $\theta_e$ is the incident angle, $\sigma$ is the standard deviation of the surface roughness, ${\uplambda}$ is the wavelength, and $\delta$ is relative permittivity. The relative properties of five materials in the Fresnel model are summarised in Table \ref{Table 11}. The relative properties are theoretical values of conventional materials in the standard environment.

\begin{table}[thp]\footnotesize
\centering
\renewcommand\arraystretch{1.5}
\caption{The relative properties of five materials.} \label{Table 11}
\addtolength{\tabcolsep}{4.8pt}
\setlength{\tabcolsep}{2pt}
\begin{tabular*}{7.95cm}{ccccccc}
	\hline
	 {Material}   &{Glass}  &{Tile}  &{Board} &{Plasterboard} &{Aluminium alloy}\\
	\hline
	  {$\delta$/ $F/m$}  &{3.5}  &{5.5}  &{2.8}  &{1.8}  &{$\infty$}\\
    \hline
	  {$\sigma$/ ${\upmu}m$}  &{0.006}  &{0.050}  &{4.800}  &{2.200}  &{4.000}\\
	\hline
\end{tabular*}
\end{table}

The angle dependence of reflection coefficients can be observed in Figs. \ref{Fig 6}a-\ref{Fig 6}e. We can find that the trends of measurement results and the Fresnel model are similar. Because of the difference between the relative parameters used in the Fresnel model and the THz measurement, there is a deviation between the results of non-metallic materials and the theoretical results. In Figs. \ref{Fig 6}a-\ref{Fig 6}d, the reflection coefficients and its growth rates of non-metallic materials increase with the incident angle. Furthermore, the growth rate increase obviously when the incident angle is above 50$\degree$. The same trends of reflection coefficients at different frequency bands can be found in \citep{r20,r17}. The reflection coefficients of metallic materials are stable at around 0.8 in Fig. \ref{Fig 6}e. And, the measurement results of aluminium alloy are lower than the theoretical results line. The reason is that the aluminium alloy may be covered with metallic oxide. The frequency dependence of reflection coefficients can also be observed in Figs. \ref{Fig 6}a-\ref{Fig 6}e. The obvious fluctuations are observed by comparing reflection coefficients along with frequency. Moreover, the fluctuations are related to frequency at the same incident angle. The fluctuations of plasterboard and board are within 0.1, which are smaller than the other three-materials fluctuations.

\subsection{Material dependence of reflection coefficients}\label{32}
To research material dependence of reflection coefficients, the average reflection coefficients of five materials are shown in Fig. \ref{Fig 6}f in the frequency from 220 to 320 GHz. The average reflection coefficients are the average of measured reflection coefficients at eight angles. Comparing the average reflection coefficients of these five materials, the average reflection coefficients of aluminium alloy are 0.28 higher than tile approximately and almost 0.5 larger than plasterboard. The aluminium alloy having the biggest complex dielectric constant reflects more power, and then in turn tile, glass, board, and plasterboard. In short, the materials with higher complex dielectric constants can reflect more power.
\section{REFLECTION COEFFICIENTS MODELING}\label{4}
In this section, a FARC model is proposed to describe the dependence of reflection coefficients on frequencies, incident angles and materials.

\begin{figure}[!t]
\centerline{\includegraphics[width=7.95cm]{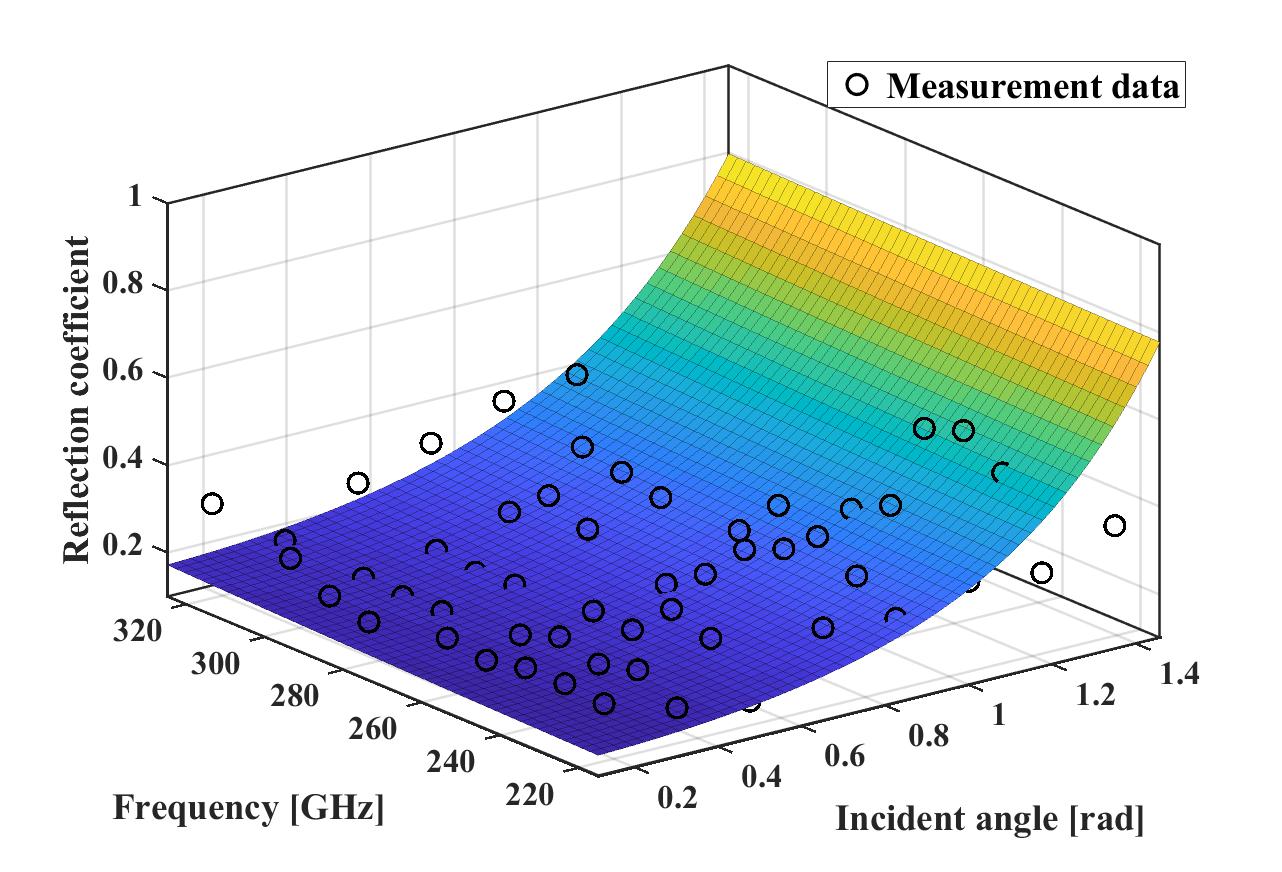}}
\caption{Reflection coefficients fitting of plasterboard.}\label{Fig 9}
\end{figure}

The FARC model is inspired by the Lorenz model, Drude model and Fresnel reflection coefficient model. Lorenz model representing the relationship between frequency and the dielectric constant of non-metallic materials is written as \citep{r100}:

\begin{equation}\label{e_14}
\begin{split}
\delta_L=1+\frac{\omega_p^2}{\omega_0^2-\omega^2-j\gamma\omega},
\end{split}
\end{equation}

\noindent with

\begin{equation}\label{e_16}
\begin{split}
\omega_p^2=\frac{N{{\rm{e}}^2}}{{\rm{m}}{\upepsilon_0}},
\end{split}
\end{equation}

\noindent where $j$ is the imaginary unit, $N$ is the number of electrons per unit volume, \rm{e} is unit positive charge, $\rm{m}$ is the mass of electron, ${\upepsilon_0}$ is permittivity of vacuum, $\gamma$ is damping constant, $\omega_0$ is resonant frequency, and $\omega$ is angular velocity. \rm{e}, $\rm{m}$, and $\epsilon_0$ are $1.6\times10^{-19}$ C, $9.3\times10^{-31}$ kg, and $8.85\times10^{-12}$ F/m, respectively. And, Drude model representing the relationship between frequency and the dielectric constant of metallic materials is expressed as \citep{r100}:

\begin{equation}\label{e_15}
\begin{split}
\delta_D=1-\frac{\omega_p^2}{\omega^2+j\gamma\omega}.
\end{split}
\end{equation}

In Eq. \ref{e_14} and Eq. \ref{e_15}, $\omega$ equals to $2{\uppi}$$f$, where $f$ is the frequency. Therefore, the Lorenz and Drude model can describe the dielectric constant continuously with frequency. To characterize the reflection coefficients in the Fresnel model at consecutive frequency, the dielectric constant of non-metallic and metallic materials should be expressed by Lorenz and Drude model, respectively. Thus, we substitute Eq. \ref{e_14} and Eq. \ref{e_15} into Equation Eq. \ref{e_10}. The FARC model of the non-metallic materials is written as:

\begin{footnotesize}
\begin{equation}\label{e_17}
\begin{split}
\Gamma_{NM}=e^{-8(\frac{{\uppi}{\upsigma}{\rm{cos}}\theta_e}{\uplambda})^2}\left(\frac{{\rm{cos}}\theta_e-\sqrt{1+\frac{\omega_p^2}{\omega_0^2-\omega^2-j\gamma\omega}-{\rm{sin}}^2\theta_e}}{{\rm{cos}}\theta_e+\sqrt{1+\frac{\omega_p^2}{\omega_0^2-\omega^2-j\gamma\omega}-{\rm{sin}}^2\theta_e}}\right),
\end{split}
\end{equation}
\end{footnotesize}

\noindent And, the FARC model of metallic materials is expressed as:

\begin{small}
\begin{equation}\label{e_18}
\begin{split}
\Gamma_{MM}=e^{-8(\frac{{\uppi}{\upsigma}{\rm{cos}}\theta_e}{\uplambda})^2}\left(\frac{{\rm{cos}}\theta_e-\sqrt{1-\frac{\omega_p^2}{\omega^2+j\gamma\omega}-{\rm{sin}}^2\theta_e}}{{\rm{cos}}\theta_e+\sqrt{1-\frac{\omega_p^2}{\omega^2+j\gamma\omega}-{\rm{sin}}^2\theta_e}}\right).
\end{split}
\end{equation}
\end{small}

The FARC model can calculate the reflection coefficients with the physical parameters of materials such as $\sigma$, $\gamma$, $\omega_0$, and $\omega_p^2$. Moreover, the FARC model describes the reflection coefficients with frequency and angle in wide frequency bands continuously with only one set of physical parameters. Compared with the two-dimensional dependence described by FARC, the Fresnel model only involves the one-dimensional of angle.

We change the $\omega$ and $\lambda$ to ${\rm{2{\uppi}}}$$f$ and ${\rm{c}}/f$, respectively, where $f$ is in GHz and ${\rm{c}}$ is the velocity of light. To characterize the model more simply, we separate the $\theta_e$ and $f$ in Eq. \ref{e_17} and Eq. \ref{e_18}. For example, Eq. \ref{e_17} can be deduced as:

\begin{tiny}
\begin{equation}\label{e_21}
\begin{split}
\Gamma_{NM}'=e^{(-\frac{8{\uppi}^2\sigma^2}{{\rm{c^2}}}f^2{\rm{cos}}^2\theta_e)}\left(\frac{{\rm{cos}}\theta_e-\sqrt{1+\frac{\frac{\omega_p^2}{{2{\uppi}}{\gamma}}}{\frac{\omega_0^2}{{2{\uppi}}{\gamma}}-\frac{2{\uppi}}{\gamma}f^2-jf}-{\rm{sin}}^2\theta_e}}{{\rm{cos}}\theta_e+\sqrt{1+\frac{\frac{\omega_p^2}{{2{\uppi}}{\gamma}}}{\frac{\omega_0^2}{{2\pi}{\gamma}}-\frac{2{\uppi}}{\gamma}f^2-jf}-{\rm{sin}}^2\theta_e}}\right).
\end{split}
\end{equation}
\end{tiny}

We set the separated physical parameters of the same materials in Eq. \ref{e_21} to constants. Moreover, because of the large magnitudes of ${8{\uppi}^2\sigma^2/\rm{c^2}}$, $\omega_p^{2}/{2{\uppi}}{\gamma}$ and $\omega_0^{2}/{2{\uppi}}{\gamma}$, we utilize powers of 10 to reduce their fitting complexity. Thus, the statistical FARC model of non-metallic materials can be written as:

\begin{scriptsize}
\begin{equation}\label{e_19}
\begin{split}
\Gamma_{NMS}=e^{(-10^af^2{\rm{cos}}^2\theta_e)}\left(\frac{{\rm{cos}}\theta_e-\sqrt{1+\frac{10^b}{10^c-df^2-jf}-{\rm{sin}}^2\theta_e}}{{\rm{cos}}\theta_e+\sqrt{1+\frac{10^b}{10^c-df^2-jf}-{\rm{sin}}^2\theta_e}}\right),
\end{split}
\end{equation}
\end{scriptsize}

\noindent where $a$, $b$, $c$, and $d$ can be derived as ${\lg({8{\uppi}^2{\sigma}^2/\rm{c^2}})}$, ${\lg(\omega_p^{2}/{2{\uppi}}{\gamma})}$, ${\lg(\omega_0^{2}/{2{\uppi}}{\gamma})}$, and ${2{\uppi}/{\gamma}}$, respectively. Therefore, $a$, $b$, $c$, and $d$ can be seen constants to be fitted statistically. Moreover, the statistical FARC model of metallic materials can be got in the same way. The statistical FARC model of metallic materials is expressed as:

\begin{footnotesize}
\begin{equation}\label{e_20}
\begin{split}
\Gamma_{MMS}=e^{(-10^af^2{\rm{cos}^2}\theta_e)}\left(\frac{{\rm{cos}}\theta_e-\sqrt{1-\frac{10^b}{df^2+jf}-{\rm{sin}}^2\theta_e}}{{\rm{cos}}\theta_e+\sqrt{1-\frac{10^b}{df^2+jf}-{\rm{sin}}^2\theta_e}}\right).
\end{split}
\end{equation}
\end{footnotesize}

To verify the feasibility of this model, we fit the five materials reflection coefficients based on the measurement results. As a representative, the results of board are shown in Fig. \ref{Fig 9}. The points of measured data fit well with the FARC model. And, most deviation values between measurement and the model results are less than 0.05. The fitted parameters of five materials are summarized in Table \ref{Table 8}. These four parameters of non-metallic materials are similar. That is because the reflection coefficients tendencies of non-metallic materials varying with angle are similar. The fitting parameters of metallic materials are much different from non-metallic materials due to its reflection coefficients different tendencies from non-metallic materials. Furthermore, the root-mean-square error (RMSE) of glass, tile, board, plasterboard, and aluminium alloy board are 0.11, 0.12, 0.10, 0.08, and 0.16, respectively. This demonstrates the well performance of the proposed FARC model. In summary, the proposed FARC model can characterize the frequency and angle well for reflection coefficients.

\begin{table}[!h]
\begin{center}
\setlength{\tabcolsep}{6pt}  
\footnotesize
\renewcommand\arraystretch{1.5}  
\caption{Fitted parameters of five materials at nine frequency points.}\label{Table 8}
\vglue8pt
\begin{tabular}{ccccccc}  
 \hline
  {Frequency}   &{$a$}  &{$b$}  &{$c$} &{$d$} &{RMSE} \\     
 \hline
   {Glass}  &{-15.45}  &{3.93}  &{3.97}  &{0.06}  &{0.11}\\
    \hline
   {Tile}  &{-15.18}  &{3.96}  &{3.72}  &{0.02}  &{0.12}\\
    \hline
   {Board}  &{-15.30}  &{3.89}  &{4.04}  &{0.03}  &{0.10}\\
    \hline
   {Plasterboard}  &{-15.66}  &{3.57}  &{4.33}  &{0.10}  &{0.08}\\
    \hline
   {Aluminium alloy}  &{-15.31}  &{6.26}  &{-}  &{0.002}  &{0.16}\\
    \hline
\end{tabular}
\end{center}
\end{table}

\section{CONCLUSION}\label{5}

This letter focuses on the reflection coefficients of building materials analysis and modeling in THz bands. Based on an extensive measurement campaign from 220 to 320 GHz, we get the reflection coefficients of five building materials. Also, the dependence of reflection coefficients on the frequencies, incident angles, and materials is investigated by comparing the measurement results and the Fresnel model. To further describe these dependencies of reflection coefficients, a FARC model and a statistical FARC model are proposed based on the Fresnel, Lorenz, and Drude model. By fitting all measured data with the statistical FARC model, the reflection coefficients of five materials are got in continuous large bands. Generally, this work is helpful for understanding THz channel propagation mechanism and simulating THz channel.
\bibliographystyle{fitee}
\bibliography{ref}
\end{document}